\documentclass[pra,twocolumn,showpacs,superscriptaddress,floatfix, reprint]{revtex4-1}

 \usepackage{latexsym}
 \usepackage{amsmath}
 \usepackage{amsfonts}
 \usepackage{graphicx}
 \usepackage{amssymb}
 \usepackage{subfigure}
 \usepackage{tabularx}
 \usepackage{color}
 \usepackage{braket}
 \usepackage[caption=false]{subfig}

\newcommand{\RNum}[1]{\uppercase\expandafter{\romannumeral #1\relax}}

\begin{document}
\title{Microwave assisted efficient four-wave mixing} 
\author{Nawaz Sarif Mallick}
\email{nawaz.phy@gmail.com}

\author{Sankar De}
\email{sankarde@gmail.com}
\affiliation{Saha Institute of Nuclear Physics, 1/AF, Bidhannagar, Kolkata 700064, India}

\date{\today}

\begin{abstract}
We theoretically investigate a $N$-type $^{87}$Rb atomic system for efficient generation and control of a non-degenerate Four wave mixing (FWM) signal in pulsed regime.
The susceptibility of the atomic medium is customized as a gain profile by a weak probe field and two strong control fields  
which allow us to generate the pulsed FWM signal. We study the propagation dynamics of the generated FWM signal inside the nonlinear medium.
The FWM signal obtains the exact shape of probe pulse and travels without changing the shape whereas the probe pulse is absorbed inside the nonlinear medium.
The conversion efficiency of this scheme without a MW field is 5.8\% which can be enhanced further by changing the control field intensity and optical depth.
However, a MW field that couples two metastable ground states enhances the conversion efficiency three times (15.55\%).   
The generation and control of such FWM signal in pulsed regime has important applications in signal processing, optical communication and quantum information science.
\end{abstract}

\maketitle
\section{Introduction} 

Four wave mixing is a well-known nonlinear phenomena which has been studied in diverse systems such as photonic crystal \cite{Andreev:02,Li:12},
optical fiber \cite{Su:22}, atomic medium \cite{Noh:21,Wu:22}, quantum dot \cite{Flayyih:13,Duan:22}. 
FWM in atomic medium attracts particular interest due to its numerous advantages over other systems.
Many theoretical and experimental studies have been performed to demonstrate the FWM process in atomic system.
Four level $N$-type \cite{PhysRevA.89.023839,Mallick_2020,Chuang:23,BackFWM}, $Y$-type \cite{Zhang:16}, double $\Lambda$-type \cite{Lee:16,deSilans:11},
diamond-type \cite{Brekke:15,Brekke:19,deMelo:14,Wang:20,PhysRevA.78.063830} atomic systems become very popular for the investigation of FWM process.
In a four level atomic system, three electromagnetic fields of frequency $\omega_1$, $\omega_2$, $\omega_3$ nonlinearly interact with the atoms
and generate FWM signal which has a frequency $\omega_g=\pm\omega_1 \pm \omega_2 \pm \omega_3$. For a non-degenerate FWM signal,
$\omega_1$, $\omega_2$, $\omega_3$ has to be different from each other \cite{Wu:22,PhysRevA.78.013834}.  
Most of the initial studies on FWM in atomic system are carried out with continuous wave (CW) lasers \cite{Wang:10,Brekke:13}.  
Later on this nonlinear process has been investigated using pulsed laser \cite{PhysRevA.89.023839,Mallick_2020} and
structured light beams \cite{Mallick:20,Swaim:18}.
Chang-Kai $et$ $al.$ experimentally demonstrates FWM using pulsed laser in a $N$-type cold atomic system \cite{PhysRevA.89.023839}.
They observe FWM conversion efficiency of 3.8\% when control laser intensity is low and 46\% when control laser intensity is high.
Note that FWM conversion efficiency can be enhanced considerably by increasing the control field intensity and optical depth of the medium
\cite{BackFWM,PhysRevA.83.053819,PhysRevA.70.061804,Hsiao:14}.
Along with the improvement of FWM conversion efficiency, the shape of the generated FWM signal at medium output becomes equally important.        
A well shaped FWM signal has many applications in signal processing and optical communications \cite{ANDREKSON2023170740,4430,Jung:06,PhysRevA.91.013843}.

\begin{figure}
\includegraphics[scale=0.3]{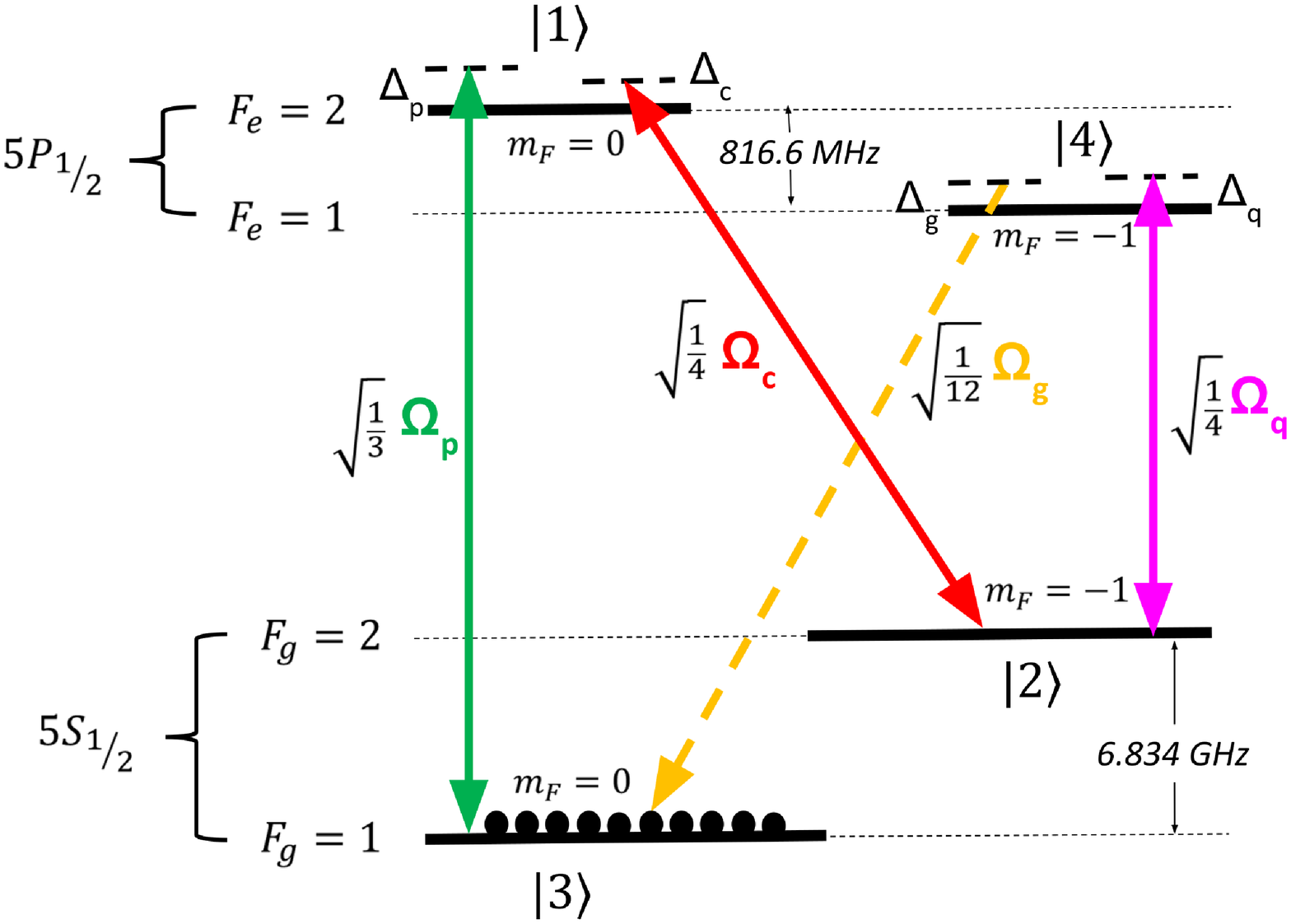}
\caption{A simple illustration of the $N$-type atomic system. The energy levels correspond to $^{87}$Rb $D_1$-line transition
($5 S_{1/2}\rightarrow 5 P_{1/2}$). The metastable ground states are deﬁned as $\ket3$=$\ket {F_g=1, m_{F}=0}$, $\ket2$=$\ket {F_g=2, m_{F}=-1}$ and
the excited states are deﬁned as $\ket1$=$\ket {F_e=2, m_{F}=0}$, $\ket4$=$\ket {F_e=1, m_{F}=-1}$. One probe field ($\Omega_p$) and two control fields
($\Omega_c$, $\Omega_q$) initiate the non-degenerate FWM signal with frequency $\omega_g=\omega_p-\omega_c+\omega_q$.
The square root terms represent the coupling strengths (Clebsch-Gordan coefﬁcient) of the corresponding transitions.}
\label{Figure1}
\end{figure}

In this work, we study a non-degenerate FWM scheme in a $N$-type atomic system as shown in Fig. \ref{Figure1}.
One probe field of frequency $\omega_p$ and two control fields of frequency $\omega_c$, $\omega_q$ resonantly interact with the $^{87}$Rb atoms and generate the FWM
signal of frequency $\omega_g=\omega_p - \omega_c +\omega_q$. For the efficient FWM generation, the phase-matching condition,
$\vec{k}_p+\vec{k}_q=\vec{k}_c+\vec{k}_g$ has to be satisfied rigorously \cite{PhysRevA.85.063821,PhysRevA.97.063806}.
We achieve this phase-matching condition by considering a collinear geometry inwhich the FWM signal generates
along the direction of probe pulse. In the present FWM scheme, the probe pulse is absorbed gradually inside the atomic medium as shown in Fig. \ref{Figure4}.
We observe that a FWM signal gradually enhances along the direction of propagation. The FWM signal obtains the exact shape of probe pulse and
propagates through the nonlinear atomic medium without changing the shape. The efficiency of the FWM process ($\eta_{eff}$) is 5.8\% which can be
enhanced further by changing the control field intensity and optical depth.
We also demonstrate how the FWM conversion efficiency can be enhanced significantly by considering an additional MW field ($\Omega_{\mu}$).
The MW field is generated using a microwave cavity which has a resonant frequency of 6.834 GHz \cite{PhysRevA.80.023820}.
When we install the Rb vapor cell inside the microwave cavity, 
a MW field of frequency 6.834 GHz couples two metastable ground states as shown in Fig. \ref{Figure5}.
Note that the dark state formed by the probe and control fields collapse
in presence of MW field and there is a MW induced population redistribution \cite{K.V.:15,PhysRevA.94.053851} in the $N$-type system which enhances the efficiency
significantly. 

The paper is organized as follows. In section \ref{THEORETICAL}, we configure the interaction of a four-level $N$-type atomic system
with electromagnetic fields using density matrix formalism. We obtain the dynamical equations for the $N$-type system
using Liouville's equation. In section \ref{Generation}, we derive the propagation equations for the probe and FWM signal using Maxwell's equations.
We study the FWM signal generation and calculate the FWM conversion efficiency. We derive an analytical expression to explain the shape of the FWM signal.
In section \ref{MW field}, we discuss how the FWM conversion efficiency can be enhanced significantly by considering an additional MW field.
Finally, in section \ref{CONCLUSION}, we conclude this work.  

\section{Model configuration}
\label{THEORETICAL}
\begin{figure}
\begin{center}
\includegraphics[scale=0.35]{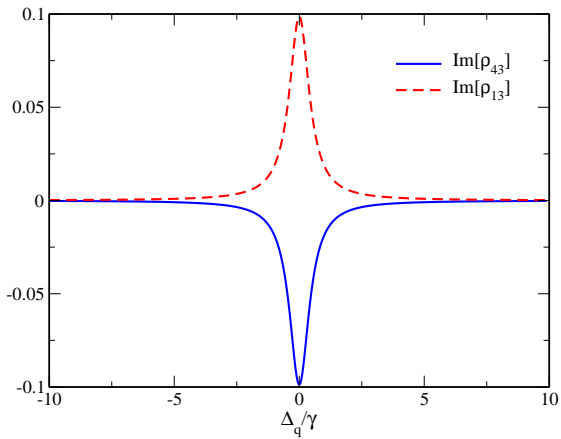}
\caption{Imaginary part of probe coherence, Im[$\rho_{13}$] and FWM signal coherence, Im[$\rho_{43}$] are plotted with control field detuning, $\Delta_q$.
Other detunings are $\Delta_p=0$, $\Delta_c=0$.}
\label{sus}
\end{center}
\end{figure}

In this section, we study a four level $N$-type atomic system for the generation and control of a non-degenerate FWM signal.
The energy levels of the $N$-type system correspond to $^{87}$Rb $D_1$-line transition ($5 S_{1/2}\rightarrow 5 P_{1/2}$).
The system interacts with three electromagnetic fields as shown in Fig. \ref{Figure1}. A weak probe field of frequency $\omega_p$
couples $\ket 1 \leftrightarrow \ket 3$ transition. Two strong control fields of frequency $\omega_c$ and $\omega_q$ couple
$\ket 1 \leftrightarrow \ket 2$ and $\ket 4 \leftrightarrow \ket 2$ transitions respectively. The probe and control fields are
defined as
\begin{equation}\label{eq:field}
\vec{E}_{j}(z,t)=\hat e_{j}\mathcal E_{0j}(z,t)e^{i(k_j z-\omega_j t)}+c.c.,	
\end{equation}
where $\hat e_{j}$ is the polarisation unit vector, $\mathcal E_{0j}(z,t)$ is the space-time dependent amplitude and $k_{j}=\omega_{j}/c$ is
the propagation constant along z-direction. The subscript, $j\in \{p,c,q\}$ represents the probe field, first control field and second control field.
Now, the interaction between the energy levels with three optical ﬁelds are described by the following interaction Hamiltonian under electric-dipole approximation
\begin{equation}\label{eq:Interaction}
\begin{aligned}
H^{'}=&\hbar \omega_{13}\ket1\bra1+\hbar(\omega_{13}-\omega_{12})\ket2\bra2\\
+&\hbar(\omega_{13}-\omega_{12}+\omega_{42})\ket4\bra4-\hbar\Omega_{p}e^{-i\omega_{p}t}\ket1\bra3\\
-&\hbar\Omega_{c}e^{-i\omega_{c}t}\ket1\bra2-\hbar\Omega_{q}e^{-i\omega_{q}t}\ket4\bra2 + h.c.,\\
\end{aligned}
\end{equation}
In Eq. \ref{eq:Interaction}, the probe and control Rabi frequencies are defined as
\begin{equation}\label{eq:Rabi}
\Omega_p=\frac{\vec{d}_{13}.\hat e_p}{\hbar} \mathcal E_{0p},~~\Omega_c=\frac{\vec{d}_{12}.\hat e_c}{\hbar} \mathcal E_{0c},
~~\Omega_q=\frac{\vec{d}_{42}.\hat e_q}{\hbar} \mathcal E_{0q}.
\end{equation}
In Eq. \ref{eq:Rabi}, $\vec{d}_{13}$, $\vec{d}_{12}$ and $\vec{d}_{42}$ are the transition dipole moments between states $\ket 1 \leftrightarrow \ket 3$,
$\ket 1 \leftrightarrow \ket 2$ and $\ket 4 \leftrightarrow \ket 2$ respectively. Next, an unitary transformation operation is performed on Eq. \ref{eq:Interaction}
to remove the time dependence from the Hamiltonian. The time independent Hamiltonian is given by
\begin{equation}\label{eq:Hamiltonian}
\begin{aligned}
H=-&\hbar\Delta_{p}\ket1\bra1-\hbar(\Delta_{p}-\Delta_{c})\ket2\bra2\\
-&\hbar(\Delta_{p}-\Delta_{c}+\Delta_{q})\ket4\bra4\\
-&\hbar\Omega_p\ket1\bra3-\hbar\Omega_c\ket1\bra2-\hbar\Omega_q\ket4\bra2 + h.c.,\\
\end{aligned}
\end{equation}
where the probe and control detunings $\Delta_{p}$, $\Delta_{c}$ and $\Delta_{q}$ are defined as
\begin{equation}\label{eq:detuning}
\Delta_{p}=\omega_p-\omega_{13}, \,\, \Delta_{c}=\omega_c-\omega_{12}, \,\, \Delta_{q}=\omega_q-\omega_{42}.
\end{equation}
The dynamics of atomic population and coherence are studied with the help of Liouville equation
\begin{equation}\label{eq:Liouville}
\dot \rho=-\frac{i}{\hbar}[H,\rho]+ \mathcal{L}_\rho ,
\end{equation}
where $\rho$ is the density operator of the system and $\mathcal{L}_\rho$ is the Lindbald operator which incorporates spontaneous decay processes
of the $N$-system. The spontaneous decay rate from the excited state $\ket 1$ to ground states $\ket 3$ and $\ket 2$ are denoted by
$\gamma_{31}$ and $\gamma_{21}$ respectively. Similarly, $\gamma_{34}$ and $\gamma_{24}$ represent the spontaneous decay rate from the excited state $\ket 4$
to the ground states $\ket 3$ and $\ket 2$ respectively. Also, the dephasing between the ground states due to collision is $\gamma_c$.
Now, we put Eq. (\ref{eq:Hamiltonian}) into the Liouville's Eq. (\ref{eq:Liouville}) and
derive the following equations of atomic population and coherence
\begin{gather}
\begin{aligned}
\dot\rho_{11}&=-(\gamma_{31}+\gamma_{21})\rho_{11}+i\Omega_p\rho_{31}+i\Omega_c\rho_{21}-i\Omega_p^*\rho_{13}-i\Omega_c^*\rho_{12},\\
\dot\rho_{12}&=[i\Delta_c-\frac{1}{2}(\gamma_{31}+\gamma_{21})]\rho_{12}+i\Omega_p\rho_{32}+i\Omega_c(\rho_{22}-\rho_{11})\\
&-i\Omega_q\rho_{14},\\
\dot\rho_{13}&=[i\Delta_p-\frac{1}{2}(\gamma_{31}+\gamma_{21})]\rho_{13}+i\Omega_p(\rho_{33}-\rho_{11})+i\Omega_c\rho_{23},\\
\dot\rho_{14}&=[i(\Delta_c-\Delta_q)-\frac{1}{2}(\gamma_{31}+\gamma_{21}+\gamma_{34}+\gamma_{24})]\rho_{14}+i\Omega_p\rho_{34}\\
&+i\Omega_c\rho_{24}-i\Omega_q^*\rho_{12},\\
\dot\rho_{22}&=\gamma_{21}\rho_{11}+\gamma_{24}\rho_{44}+i\Omega_c^*\rho_{12}+i\Omega_q^*\rho_{42}-i\Omega_c\rho_{21}-i\Omega_q\rho_{24},\\
\dot\rho_{23}&=[i(\Delta_p-\Delta_c)-\gamma_c]\rho_{23}+i\Omega_c^*\rho_{13}+i\Omega_q^*\rho_{43}-i\Omega_p\rho_{21},\\
\dot\rho_{24}&=-[i\Delta_q+\frac{1}{2}(\gamma_{34}+\gamma_{24})]\rho_{24}+i\Omega_c^*\rho_{14}+i\Omega_q^*(\rho_{44}-\rho_{22}),\\
\dot\rho_{33}&=\gamma_{31}\rho_{11}+\gamma_{34}\rho_{44}+i\Omega_p^*\rho_{13}-i\Omega_p\rho_{31},\\
\dot\rho_{43}&=[i(\Delta_p-\Delta_c+\Delta_q)-\frac{1}{2}(\gamma_{34}+\gamma_{24})]\rho_{43}-i\Omega_p\rho_{41}+i\Omega_q\rho_{23},\\
\dot\rho_{44}&=-(\dot\rho_{11}+\dot\rho_{22}+\dot\rho_{33}),\\
\dot\rho_{ij}&=\dot\rho_{ji}^*,
\end{aligned}
\label{eq:r}
\raisetag{15pt}
\end{gather}
where the overdot denotes for the time derivative and the star (*) represents the complex conjugate.
In Eq. \ref{eq:r}, diagonal density matrix elements $\rho_{11}$, $\rho_{22}$, $\rho_{33}$ and $\rho_{44}$ satisfy the conservation of population i.e.
$\rho_{11}$ + $\rho_{22}$ + $\rho_{33}$ + $\rho_{44}$ = 1. Note that off-diagonal density matrix elements $\rho_{13}$ and $\rho_{43}$ control the propagation
dynamics of the probe pulse ($\Omega_p$) and generated FWM signal ($\Omega_g$) respectively. In order to obtain atomic coherence $\rho_{13}$ and $\rho_{43}$,
we solve all coupled equations (Eq. \ref{eq:r}) numerically under steady state condition. The imaginary part of probe coherence, Im[$\rho_{13}$] and
FWM signal coherence, Im[$\rho_{43}$] are shown in Fig. \ref{sus}.
Solid blue curve (Im[$\rho_{43}$]) in Fig. \ref{sus} shows negative absorption (gain) which indicates that a FWM signal generation is possible along
$\ket 4 \rightarrow\ket 3$ transition. Similarly, dashed red curve (Im[$\rho_{13}$]) shows positive absorption which means the probe pulse is absorbed inside
the atomic medium.     

\section{Generation of non-degenerate FWM signal}
\label{Generation}
\begin{figure}
\begin{center}
\includegraphics[scale=0.7]{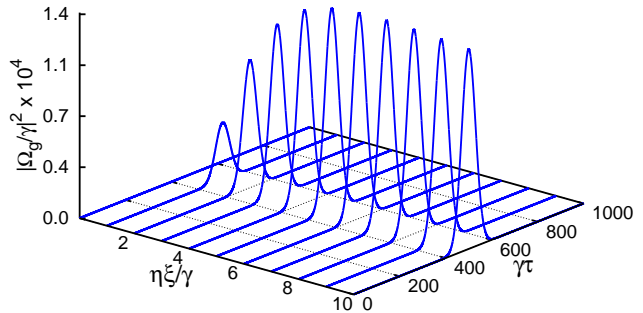}
\caption{FWM signal generation as a function of position and time. The parameters are : $\Omega^0_p=0.05\gamma$, $\Omega^0_c=5.0\gamma$, $\Omega^0_q=5.0\gamma$,
$\Delta_p=0$, $\Delta_c=0$, $\Delta_q=0$, $\gamma_c \simeq 1\times 10^3$ Hz, $\lambda=795 \times 10^{-7}$ cm, $\mathcal N=1.3\times 10^9$ atoms/cm$^3$,
$\sigma_p=60/\gamma$, $\gamma \tau_0=500$.}
\label{Figure3}
\end{center}
\end{figure}
In this section, we take advantage of Maxwell's equations to study spatio-temporal evaluation of the FWM signal along with probe pulse through the
nonlinear atomic medium. We can write the wave equation for the probe, control and FWM fields as
\begin{equation}\label{eq:wave_eq}
\left(\nabla^2+\frac{1}{c^2}\frac{\partial^2}{\partial t^2}\right)\vec{E}=\frac{4\pi}{c^2}\frac{\partial^2 \vec{\mathcal P}}{\partial t^2},
\end{equation} 
where $\vec{\mathcal P}$ is the total polarization induced by the total electric field $\vec{E}=\vec{E}_p+\vec{E}_c+\vec{E}_q+\vec{E}_g$. Note that
$\vec{\mathcal P}$ is the source term of Eq. \ref{eq:wave_eq} and it controls the linear and non-linear phenomena inside the atomic medium.
We can write $\vec{\mathcal P}$ in terms of atomic density ($\mathcal{N}$) and atomic coherences as
\begin{equation}\label{eq:polarization}
\begin{aligned}
\vec{\mathcal P}&=\mathcal N(\vec{d}_{13}\rho_{13}e^{-i\omega_p t}+\vec{d}_{12}\rho_{12}e^{-i\omega_c t}+\vec{d}_{42}\rho_{42}e^{-i\omega_q t}\\
&+\vec{d}_{43}\rho_{43}e^{-i\omega_g t}+c.c.),
\end{aligned}
\end{equation}
Now, we insert Eq. \ref{eq:polarization} into Eq. \ref{eq:wave_eq} and rewrite the wave equation under slowly-varying envelope approximation (SVEA) as
\begin{equation}\label{eq:14}
\begin{aligned}
&\left(\frac{\partial }{\partial z}+\frac{1}{c}\frac{\partial }{\partial t}\right)\Omega_p=i\eta_p\rho_{13}(z,t),\\
&\left(\frac{\partial }{\partial z}+\frac{1}{c}\frac{\partial }{\partial t}\right)\Omega_c=i\eta_c\rho_{12}(z,t),\\
&\left(\frac{\partial }{\partial z}+\frac{1}{c}\frac{\partial }{\partial t}\right)\Omega_q=i\eta_q\rho_{42}(z,t),\\
&\left(\frac{\partial }{\partial z}+\frac{1}{c}\frac{\partial }{\partial t}\right)\Omega_g=i\eta_g\rho_{43}(z,t).
\end{aligned}
\end{equation}
We can safely neglect the propagation of the control ﬁelds ($\Omega_c$ and $\Omega_q$) because they possess much higher intensity than
the probe and FWM signal intensity \cite{PhysRevA.94.053851}. In Eq. \ref{eq:14}, $\eta_p$ and $\eta_g$ are the coupling constant for the probe and FWM signal.
We can rewrite $\eta_p$ and $\eta_g$ in terms of reduced coupling constant, $\eta=2\pi k \mathcal N |\vec{d}^{R}_{ij}|^{2}/\hbar$ as
\begin{equation}
\eta_{p}=\frac{\eta}{3} \quad;\quad \eta_{g}=\frac{\eta}{12}
\end{equation}  
where $|\vec{d}^{R}_{ij}|$ is the reduced matrix element.
Note that different coupling strength from the excited states into the ground states also revise the spontaneous decay rate
and can be written as
\begin{equation}
\gamma_{31}=\frac{\gamma}{3};\quad \gamma_{21}=\frac{\gamma}{4};\quad \gamma_{24}=\frac{\gamma}{4};\quad
\gamma_{34}=\frac{\gamma}{12}
\end{equation}   
where $\gamma=4|\vec{d}^{R}_{ij}|^{2} k^{3}/3 \hbar$ is the reduced spontaneous decay rate.

\begin{figure}
\begin{center}
\includegraphics[scale=0.4]{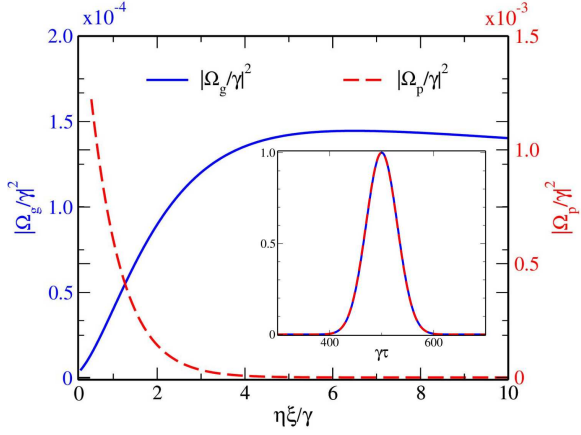}
\caption{FWM signal (solid blue curve) and probe signal (dashed red curve) are plotted along the propagation direction. Inset figure compares normalised FWM signal
(solid blue curve) at medium output and probe signal (dashed red curve) at medium input. All other parameters are the same as in ﬁgure \ref{Figure3}.}
\label{Figure4}
\end{center}
\end{figure}
Now, we incorporate a co-moving coordinate system which allow us to perform numerical computation
\begin{equation}
\tau=t-\frac{z}{c},\hspace{0.25cm} \xi=z.
\end{equation}
Therefore, in moving coordinate system, the expression $[\partial /\partial z+(1/c)\partial/\partial t]$ in Eq.(\ref{eq:14}) is replaced by $\partial/\partial \xi$.
The simultaneous solution of Eq. \ref{eq:r} and Eq. \ref{eq:14} inspect the dynamical progression of FWM signal inside the medium.
We solve the coupled partial differential equations using the Cash Karp Runge Kutta method. We start with a Gaussian shaped probe pulse whose
time dependent envelope at medium input is given by
\begin{equation}\label{eq:probeshape}
\Omega_p(\xi=0,\tau)=\Omega_p^0 e^{-(\frac{\tau-\tau_0}{\sigma_p})^2},
\end{equation}       
In Eq. \ref{eq:probeshape}, $\Omega_p^0$ is the amplitude, $\sigma_p$ is the temporal width and $\tau_0$ is the peak location.
The control fields ($\Omega_c$ and $\Omega_q$) are taken as continuous wave (cw) field. The spatio-temporal variation of generated FWM signal
is shown in Fig. \ref{Figure3}. The intensity of FWM signal raises gradually as it propagates through the atomic medium.
In Fig. \ref{Figure4}, we plot the peak intensity of FWM signal and probe pulse as a function of propagation distance.
It is clear from Fig. \ref{Figure4} that the probe pulse intensity (dashed red curve) decreases gradually as it propagates through the atomic medium.           
However, the FWM signal intensity (solid blue curve) increases gradually along the propagation direction. The inset of Fig. \ref{Figure4}
compares the shape of FWM signal and probe pulse. We plot the normalised shape of FWM signal (solid blue curve) at medium output ($\xi$=10)
and normalised shape of probe pulse (dashed red curve) at medium input ($\xi$=0). The inset of Fig. \ref{Figure4} clearly shows that
the generated FWM signal has the same Gaussian shape as the probe pulse.

\begin{figure}
\begin{center}
\includegraphics[scale=0.3]{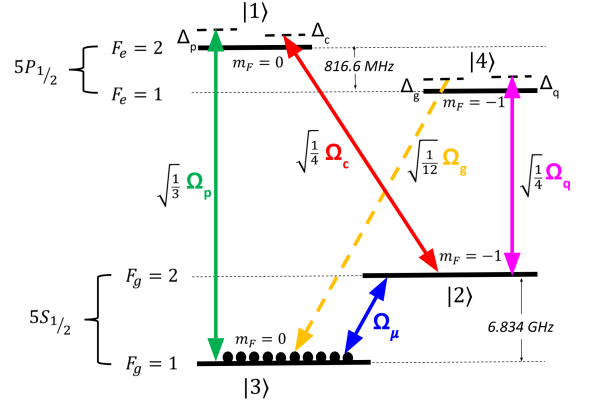}
\caption{A simple illustration of the model system inwhich an additional MW field ($\Omega_{\mu}$) couples two meta-stable ground states $\ket 3$ and $\ket 2$.}
\label{Figure5}
\end{center}
\end{figure}
Next, we require an expression of atomic coherence, $\rho_{43}$ to explain the FWM signal generation inside the nonlinear medium.  
We derive an analytical expression of $\rho_{43}$ under the weak probe approximation ($\Omega_p$ $<$ $\Omega_c$, $\Omega_q$). The approximation is
valid for all orders of control field Rabi frequencies ($\Omega_c$, $\Omega_q$) and $1^{st}$ order of probe Rabi frequency ($\Omega_p$). Under this
approximation, we can write the solution of the density matrix equations as
\begin{eqnarray}
\rho_{ij}&=\rho_{ij}^{(0)}+\Omega_p\rho_{ij}^{(1)}+\Omega_p^* \rho_{ij}^{(2)},
\end{eqnarray}
where $\rho_{ij}^{(0)}$ is the solution in the absence of $\Omega_p$ and $\rho_{ij}^{(k)}$, $k\in \{1,2\}$ is
the higher order solution in the presence of $\Omega_p$.
The steady-state value of the atomic coherence $\rho_{43}$ can be expressed by the following expression
\begin{gather}\label{rho43}
\begin{aligned}
\rho_{43}=\frac{i\Omega_p\Omega_c^*\Omega_q}{\Gamma_{13}\Gamma_{23}\Gamma_{43}[1+\frac{|\Omega_c|^2}{\Gamma_{13}\Gamma_{23}}+\frac{|\Omega_q|^2}{\Gamma_{23}\Gamma_{43}}]}
\end{aligned}
\end{gather}
where,
\begin{gather}
\begin{aligned}
\Gamma_{13}&=i\Delta_p-\frac{1}{2}(\gamma_{31}+\gamma_{21}),\\
\Gamma_{23}&=i(\Delta_p-\Delta_c)-\gamma_c,\\
\Gamma_{43}&=i(\Delta_p-\Delta_c+\Delta_q)-\frac{1}{2}(\gamma_{34}+\gamma_{24}).\nonumber
\end{aligned}
\end{gather}
Note that the atomic coherence $\rho_{43}$ generates the temporal shape of the FWM signal. The solution of the FWM signal can
be obtained from the following propagation equation
\begin{equation}\label{sol4}
\frac{\partial \Omega_g}{\partial \xi}=\frac{i\eta}{12}\left(\frac{i\Omega_p\Omega_c^*\Omega_q}{\Gamma_{13}\Gamma_{23}\Gamma_{43}[1+\frac{|\Omega_c|^2}{\Gamma_{13}\Gamma_{23}}+\frac{|\Omega_q|^2}{\Gamma_{23}\Gamma_{43}}]}\right)
\end{equation}
Eq. \ref{sol4} implies that the envelope of the FWM signal, $\Omega_{g}\propto \Omega_{p}$, in the presence of a CW control fields.
The probe pulse is considered as $\Omega_{p}=|\Omega_p^{0}| e^{-\tau^{2}/\sigma^{2}_{p}}$. Therefore, the shape of the FWM signal becomes
$\Omega_{g}\propto|\Omega_p^{0}| e^{-\tau^{2}/\sigma^{2}_{p}}$ which clearly manifests that
the temporal width of the FWM pulse is $\sigma_g=\sigma_p$.

The efficiency of the FWM process, $\eta_{eff}$ is the ratio of energy of the output generated FWM signal and energy of the input probe pulse
\cite{PhysRevA.70.053818}
\begin{equation}\label{efficiency}
\eta_{eff}=\frac{\int_{-\infty}^{\infty}|\vec{E}_g (z=L,\tau)|^2~d\tau}{\int_{-\infty}^{\infty} |\vec{E}_{p} (z=0,\tau)|^2~d\tau}
\end{equation}
In absence of MW field, the efficiency of the FWM process ($\eta_{eff}$) is 5.8\%. Note that FWM generation also depends on the intensity of control fields
$|\Omega_c|^2$, $|\Omega_q|^2$ as shown in Eq. \ref {rho43} and can be enhanced further by adjusting the control fields.

\begin{figure}
\begin{center}
\includegraphics[scale=0.7]{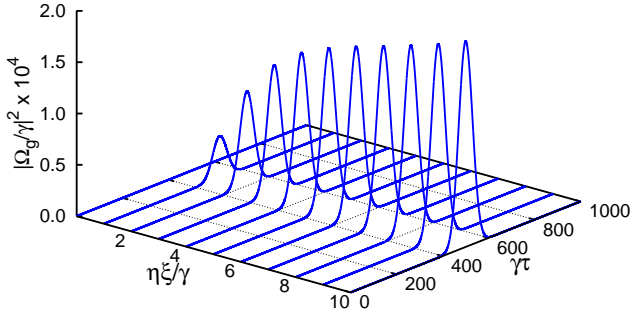}
\caption{FWM signal generation in presence of MW field as a function of position and time. The parameters are : $\Omega^0_p=0.05\gamma$, $\Omega^0_c=5.0\gamma$,
$\Omega^0_q=5.0\gamma$, $\Omega^0_{\mu}=0.05\gamma$, $\Delta_p=0$, $\Delta_c=0$, $\Delta_q=0$, $\gamma_c \simeq 1\times 10^3$ Hz, $\lambda=795 \times 10^{-7}$ cm, $\mathcal N=1.3\times 10^9$ atoms/cm$^3$,
$\sigma_p=60/\gamma$, $\gamma \tau_0=500$..}
\label{Figure6}
\end{center}
\end{figure}
\section{Effect of MW field}
\label{MW field}
In this section, we discuss how the efficiency of non-linear FWM process can be enhanced significantly by considering an additional MW field ($\Omega_{\mu}$).
The MW field couples two metastable ground states of $5 S_{1/2}$ $i.e.$ $\ket3$=$\ket {F_g=1, m_{F}=0}$, $\ket2$=$\ket {F_g=2, m_{F}=-1}$.
A simple illustration of the model system is shown in Fig. \ref{Figure5}. The frequency separation between state $\ket3$ and $\ket2$ is 6.834 GHz.
Now, we consider four fields ($\Omega_{p}$, $\Omega_{c}$, $\Omega_{q}$, $\Omega_{\mu}$) and derive 16 density matrix equations using Eq. \ref{eq:Liouville}.
The MW field ($\Omega_{\mu}$) and control fields ($\Omega_{c}$ and $\Omega_{q}$) are taken as continuous wave (cw) field.
The probe field is chosen to be a Gaussian shaped pulse which is expressed by Eq. \ref{eq:probeshape}.
Next, we solve the coupled wave equations for probe and FWM signal to understand the effect of MW field on FWM generation.

\begin{figure}
\begin{center}
\includegraphics[scale=0.33]{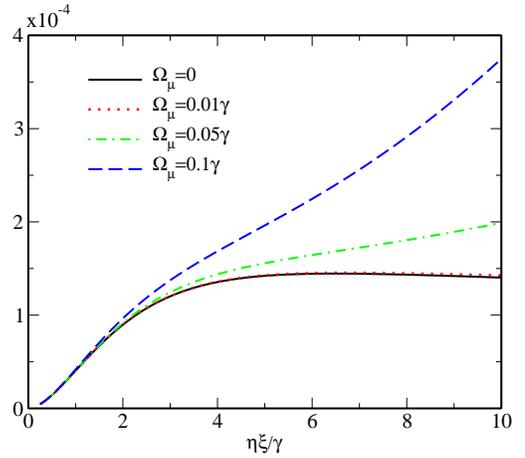}
\caption{FWM signal is plotted along the propagation direction at different values of MW field such as $\Omega_{\mu}=0$ (solid black curve),
$\Omega_{\mu}=0.01\gamma$ (dotted red curve), $\Omega^0_{\mu}=0.05\gamma$ (dash dot green curve), $\Omega^0_{\mu}=0.1\gamma$ (dashed blue).
All other parameters are the same as in ﬁgure \ref{Figure6}.}
\label{Figure7}
\end{center}
\end{figure}
The space-time dependent variation of generated FWM signal in presence of MW field is shown in Fig. \ref{Figure6}.
The FWM signal intensity is zero at medium input ($\xi$=0). The signal increases in a gradual way along the direction of propagation.
In contrast to the result presented in Fig. \ref{Figure3}, the FWM signal keep on increasing along the propagation direction in presence of requisite MW field.
It is clear from Fig. \ref{Figure6} that the shape of the generated FWM signal remains same as the probe pulse in presence of MW field.     

In Fig. \ref{Figure7}, we plot the peak intensity of the generated FWM signal as a function of propagation distance for different values of MW field.
Solid black curve in Fig. \ref{Figure7} represents FWM signal in absence of MW field ($\Omega_{\mu}=0$). Dotted red curve in Fig. \ref{Figure7} displays
the FWM signal in presence of weak MW field ($\Omega_{\mu}<\Omega_{p}$). The effect of weak MW field ($\Omega_{\mu}=0.01\gamma$) on FWM generation is
almost negligible as shown in Fig. \ref{Figure7}. In presence of requisite MW field ($\Omega_{\mu}=\Omega_{p}=0.05\gamma$),
the FWM signal generation enhances adequately which is indicated by the dash dot green curve in Fig. \ref{Figure7}. In presence of
strong ($\Omega_{\mu}>\Omega_{p}$) MW field ($\Omega_{\mu}=0.1\gamma$), there is significant enhancement of FWM generation as shown
with dashed blue curve in Fig. \ref{Figure7}. Next, we calculate the efficiency ($\eta_{eff}$) of FWM generation for three different values of MW field
using Eq. \ref{efficiency} and the outcome is presented in the following table :\\

\begin{tabularx}{0.44\textwidth} { 
  | >{\raggedright\arraybackslash}X 
  | >{\centering\arraybackslash}X 
  | >{\raggedleft\arraybackslash}X | }
 \hline
 Probe field ($\Omega_p$) & MW field ($\Omega_{\mu}$) & FWM efficiency\\
 \hline
 $0.05\gamma$  & $0.01\gamma$  & 5.90 \%  \\
\hline
 $0.05\gamma$  & $0.05\gamma$  & 8.24 \%  \\
\hline
 $0.05\gamma$  & $0.1\gamma$  & 15.55 \%  \\
\hline
\end{tabularx}\\
  
The enhancement of FWM conversion efficiency in presence of MW field can be understood with the concept of dark state.
In absence of MW field, the dark state is formed by the probe and control fields \cite{Preethi_2011,PhysRevA.80.023820}. Under this condition,
the populations remain trapped in the ground state $\ket 3$.
When the MW field couples two meta-stable ground states, dark state formation collapses and there is a MW induced population redistribution
\cite{K.V.:15,PhysRevA.94.053851,PhysRevA.80.023820} in the $N$-type system which enhances the FWM efficiency.

\section{CONCLUSION}
\label{CONCLUSION}
In conclusion, we investigate the generation of a non-degenerate FWM signal in a $N$-type $^{87}$Rb atomic system.
The susceptibility of the atomic medium is customized as a gain proﬁle by a weak probe field and two strong control fields  
which allow us to generate the FWM signal. We study the propagation dynamics of the generated FWM signal through the nonlinear medium.
The FWM signal obtains the exact shape of probe pulse and travels through the atomic medium without changing the
shape. The FWM conversion efficiency is 5.8\% which can be
enhanced further by changing the control field intensity and optical depth. Apart from this
we also demonstrate how the FWM conversion efficiency can be enhanced three times using a MW field. 
This FWM scheme has important applications in diverse fields such as signal processing, optical communication and quantum
information science \cite{PhysRevA.91.013843,Wang:17,PhysRevA.95.051802}. 

\section*{Acknowledgments}
N.S.M. and S.D. acknowledge funding from the National Mission in Interdisciplinary Cyber-Physical systems from
the Department of Science and Technology through the I-HUB Quantum Technology Foundation (Grant No. I-HUB/PDF/2021-22/007).

\bibliography{reference}
\end{document}